# Discovery of a Robust Non-Janus Hybrid MoSH Monolayer as a Two-Gap Superconductor via High-Throughput Computational Screening

Zhijing Huang[1,*], Hongmei Xie[2,*], Zhibin Gao[1], Longyuzhi Xu[2], Lin Zhang[2], Li Yang[1,†], Zonglin Gu[2,†], Shuming Zeng[2,†]

[1]College of Physical Science and Technology, Guangxi Normal University, Guilin, Guangxi 541004, China

[2]College of Physical Science and Technology, Yangzhou University, Yangzhou, Jiangsu 225009, China

*These authors contribute equally to this work

†Corresponding authors: zengsm@yzu.edu.cn (Shuming Zeng), guzonglin@yzu.edu.cn (Zonglin Gu), yangli@mailbox.gxnu.edu.cn (Li Yang).

## Abstract

The atomic-scale determination of hydrogen positions in MoSH monolayers remains experimentally challenging, and existing studies are confined to Janus-type configurations. Here, we combine high-throughput structural screening with first-principles calculations to predict a novel non-Janus Hybrid 1T'-MoSH monolayer, which energetically surpasses all previously reported MoSH phases with a binding energy of −3.02 eV. This structure emerges as a hybrid of $MoS_2$ and $MoH_2$, featuring alternating S and H atoms on both sides of the Mo layer. Comprehensive stability analyses confirm its robustness in energy, mechanics, dynamics, and thermodynamics (stable up to 1600 K). Remarkably, anisotropic Migdal-Eliashberg theory predicts Hybrid 1T'-MoSH as a two-gap superconductor with a critical temperature $T_c$ of 16.34 K, driven by strong electron-phonon coupling ($\lambda$=1.39). Substituting Mo with Hf, Ta, or Ti drastically suppresses $T_c \sim$ (0.53–2.42 K), highlighting Mo's unique role in enhancing superconductivity. Our work not only expands the family of 2D transition metal chalcogenides but also proposes a promising candidate for quantum technologies, bridging theoretical design to functional material discovery.

## 1. Introduction

Surface and interface science serves as the fundamental scientific underpinning for diverse fields such as physics[1-3], chemistry[4-6], and biology[7-9]. It encompasses materials, energy, and information-related fundamental disciplines and finds extensive applications in catalysis[10], superconductivity[11-12], and chip technologies[13]. Among numerous interfacial systems, two-dimensional (2D) interface structures have attracted significant attention from the scientific community. Their atomic-scale thickness[14-15], large surface area[16-17], wide variety of space groups[18], and unique electronic band structures[19-20] render them ideal for exploring quantum phenomena. In recent years, researchers have increasingly focused on the design, prediction, and regulation of 2D interface microstructures[21-23]. Precise theoretical design and prediction of 2D interface atomic structures, along with the construction of effective physical models, are pivotal steps for the rapid experimental synthesis of interface structures and the accurate regulation of interface properties. However, due to the complexity and diversity of atomic coordination, interface structures often exhibit variations in components, distributions, and defects. Their impacts on interface structure stability and interfacial interactions are complex and have not been comprehensively understood. As a result, accurately synthesizing, measuring, and regulating interface structures through experiments remains extremely challenging. Therefore, theoretical investigations into the design and prediction of 2D interfacial microstructures and their associated physical properties are crucial for the development of functional materials with tailored physical characteristics.

Janus transition-metal dichalcogenides (TMDs) are a class of composite 2D materials. Their structure, characterized by different chalcogen atoms covalently bonded to the two surfaces of the transition metal, exhibits distinct "sandwich" features[24-25]. In 2017, Janus MoSSe was successfully synthesized[26], which led to a leapfrog development of Janus transition metal chalcogenide materials. These materials[27-33] include MoSO, MoSTe, WSSe, SnSSe, NbSSe, PtSSe, ZrSSe, WSO, TiSO, ZrSO, HfSO, WSeTe, etc. Notable progress has been achieved in the synthesis and characterization of Janus MoSH monolayers. Pioneering work by Wan[34] et al. demonstrated the controlled synthesis of Janus MoSH via hydrogen plasma method. In this process, selective replacement of sulfur atoms in MoS₂ with hydrogen created an S-Mo-H configuration with intrinsic vertical dipoles and metallic behavior. Subsequent studies revealed that Janus MoSH phases (2H, 1T, and 1T') possess distinct structural characteristics and electronic properties, including temperature-dependent structural phase transitions[35], different proportions of H atoms[36], high carrier densities[37], piezoelectric[38] and superconducting properties[39]. However, the lightweight nature of H atoms[40] makes it extremely challenging to accurately determine the physical model of the MoSH monolayer through experimental methods, even with the most advanced High-Resolution Transmission Electron Microscopy[34]. This means that the physical model of MoSH remains ambiguous. It is worth noting that all existing studies on the structure of MoSH at present belong to the Janus structure type, and whether the MoSH monolayer with a non-Janus structure exists remains unknown.

In this work, we designed a novel Hybrid 1T'-MoSH monolayer, where S and H atoms are arranged in an alternating pattern on both sides of the Mo layer. Unlike the conventional Janus MoSH monolayer, the Hybrid 1T'-MoSH exhibits a non-Janus structure. Through a comprehensive stability analysis, we found that it exhibits high stability in terms of energy, mechanics, dynamics, and thermodynamics. Notably, it is predicted to be a two-gap superconductor with a superconducting transition temperature $T_c$ of 16.34 K. Furthermore, we revealed that substituting Mo atoms with Hf, Ta, or Ti atoms in the Hybrid 1T'-MoSH leads to a drastic reduction in $T_c$ values, ranging from 0.53 to 2.42 K. As far as we know, this study is the first to theoretically propose a physical model of the non-Janus MoSH monolayer.

## 2. Results and Discussion

To identify new and stable 2D MoSH monolayers, we combined high-throughput methods based on random method, group and graph theory (RG²) with first-principles calculations to explore the interfacial structure of MoSH monolayers. **Figure 1(a)** illustrates the flowchart for discovering a new phase structure of MoSH monolayers. The flowchart consists of three steps: (I) The RG² code randomly generated 2D interface structures with a Mo:S:H ratio of 1:1:1, which provided a dataset of initial MoSH monolayer structures; (II) First-principles calculations evaluated the energy stability of the

MoSH structures in the dataset, selecting the lowest energy structure and further assessing its dynamical, thermodynamical, and mechanical stability; (III) The newly stable MoSH monolayer was identified.

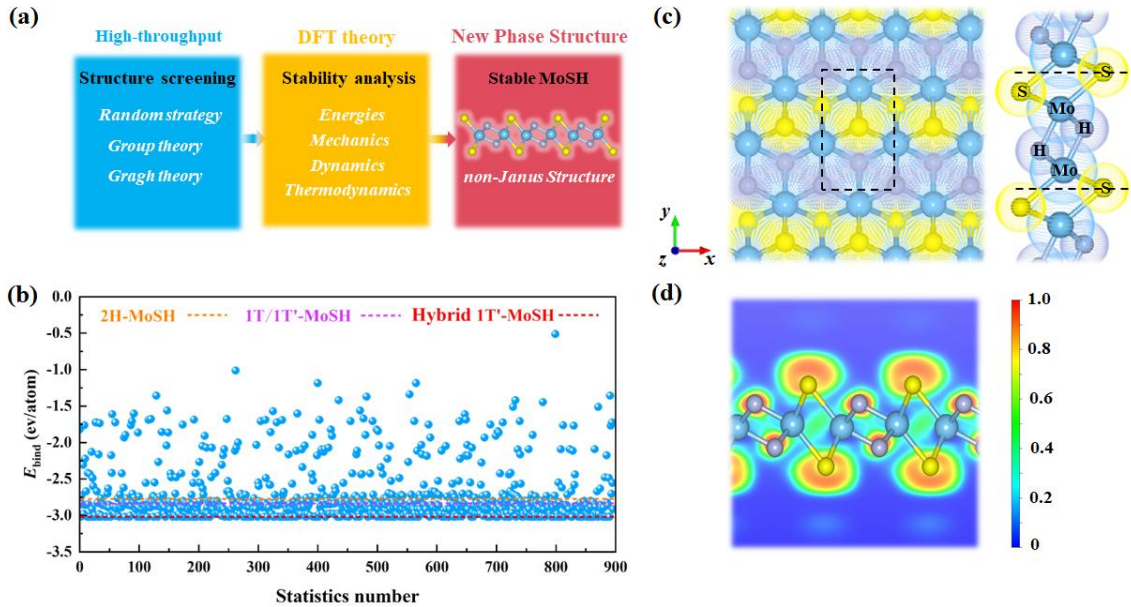

**Figure 1.** (a) Schematic illustration of the fundamental procedures for the discovery of novel stable non-Janus MoSH monolayers, (b) The binding energies ($E_{bind}$) of 896 predicted structures of MoSH monolayers, (c) Top and side views of the Hybrid 1T'-MoSH structure, (d) Electron Localization Function (ELF) for the Hybrid 1T'-MoSH monolayers.

In step I, the RG$^2$ code generated a total of 896 initial 2D structures of MoSH monolayers. In step II, the binding energy ($E_{bind}$) of the 896 MoSH monolayer structures was first calculated using the formula: $E_{bind}=(E_{total}-NE_{Mo}-NE_{S}-NE_{H_2}/2)/3N$, where $E_{total}$ represents the total energy of MoSH, $E_{Mo}$, $E_S$ are the energies of isolated Mo, S atoms, $E_{H_2}$ is the energy of a hydrogen molecule, $N$ denotes the number of Mo, S, and H atoms. A larger negative value of $E_{bind}$ indicates a higher level of energy stability for MoSH monolayer structures. As shown in **Figure 1(b)**, the $E_{bind}$ of the generated structures varies widely, with the majority clustered within a stable range of -3.1 eV to -2.5 eV. Among these discovered candidates, we identified Janus phase structures of 2H-MoSH, 1T-MoSH, and

1T'-MoSH, with $E_{bind}$ of -2.77 eV (orange dashed line), -2.83 eV (purple dashed line), and -2.84 eV (purple dashed line), respectively. These Janus phase structures are consistent with previous studies[35],[37],[39]. Interestingly, we discovered a new phase of MoSH with the lowest $E_{bind}$ of -3.02 eV (red dashed line) compared to the other candidates, suggesting that it has the highest energy stability among all the discovered candidates.

**Figure 1(c)** presents the atomic structure of the newly identified MoSH structure, which can be regarded as a hybrid compound formed through the mutual hybridization of $MoS_2$ and $MoH_2$. From both the top and side views, the structure closely resembles that of the Janus 1T'-MoSH phase. Notably, the S and H atoms in this structure are periodically distributed on both sides of the Mo atomic layer, resulting in a non-Janus configuration. This distinct arrangement of Mo, S, and H atoms differentiates it from the Janus 1T'-MoSH phase. Therefore, we define this structure as the Hybrid 1T'-MoSH monolayer in this work.

To further elucidate the structural characteristics of the Hybrid 1T'-MoSH monolayer, **Figure 1(d)** and **Table S1** provide a detailed analysis, including the 2D electron localization function (ELF), lattice constants, bond lengths and Bader charge. As shown in **Figure 1(d)**, the Mo-H and Mo-S bonds in the Hybrid 1T'-MoSH monolayer primarily exhibit ionic bonding characteristics, similar to those observed in the 2H-, 1T- and 1T'-MoSH phases[35]. **Table S1** shows the structural differences of MoSH in different phases. Compared to the Janus 2H-MoSH, 1T-MoSH, and 1T'-MoSH phases, Hybrid 1T'-MoSH exhibits a smaller lattice constant and Mo-H bond length ($d_1$ = 1.91 Å), as well as a greater negative charge on the sulfur atoms.

**Table 1.** Elastic constants of 2H, 1T, 1T' and Hybrid 1T' phases of Janus MoSH monolayer, in GPa.

| MoSH | $C_{11}$ | $C_{12}$ | $C_{22}$ | $C_{66}$ |
|---|---|---|---|---|
| 2H[35] | 78.25 | 22.21 | 71.87 | -4.36 |
| 1T[35] | 66.15 | 44.75 | 58.56 | 2.18 |
| 1T'[35] | 32.13 | 23.73 | 37.02 | 1.97 |
| Hybrid-1T' | 65.24 | 30.95 | 58.25 | 0.77 |

Furthermore, the mechanical stability of Hybrid 1T'-MoSH is estimated by calculating the elastic constants. The mechanically stable crystal of MoSH should satisfy the criterion $C_{11}C_{22} > C_{12}C_{12}$ and $C_{66} > 0$. In previous study[35], the elastic constants have already shown that the mechanical properties of the 1T, and 1T' phases are stable. However, it is worth noting that the mechanical stability of the 2H phase is unfavorable due to $C_{66} < 0$. Similar to the 1T and 1T' phases, the mechanical stability of the Hybrid 1T'-MoSH also satisfies the lattice stability criteria, indicating that the Hybrid 1T'-MoSH is mechanically stable. The corresponding values are presented in **Table 1**.

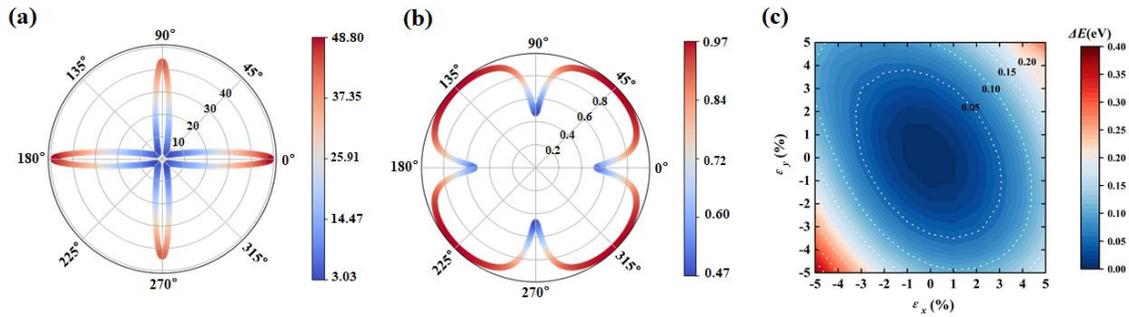

**Figure 2.** (a) Young's modulus, (b) Poisson's ratio, and (c) Surface plot of strain energy ($\Delta E$) corresponding to in-plane strain ($\varepsilon_x$, $\varepsilon_y$) for Hybrid 1T'-MoSH.

To further investigate the mechanical properties of the Hybrid 1T'-MoSH, the Young's modulus ($Y(\theta)$) and Poisson's ratio ($v(\theta)$) of the Hybrid 1T'-MoSH structure were calculated using the following equations:

$$Y(\theta) = \frac{C_{11}C_{22} - C^2_{12}}{C_{22}(\cos\theta)^4 + A(\cos\theta)^2(\sin\theta)^2 + C_{11}(\sin\theta)^4}$$

$$v(\theta) = \frac{C_{12}(\cos\theta)^4 - B(\cos\theta)^2(\sin\theta)^2 + C_{12}(\sin\theta)^4}{C_{22}(\cos\theta)^4 + A(\cos\theta)^2(\sin\theta)^2 + C_{11}(\sin\theta)^4}$$

where $A = \frac{C_{11}C_{22} - C^2_{12}}{C_{66}} - 2C_{12}$, $B = C_{11} + C_{22} - \frac{C_{11}C_{22} - C^2_{12}}{C_{66}}$, and $\theta$ is the angle to the axes of 100.

As shown in **Figure 2(a)**, the $Y(\theta)$ exhibits distinct anisotropic characteristics for Hybrid 1T'-MoSH. Notably, the $Y(\theta)$ value of 3.03~48.80 N·m⁻¹ is slightly higher than those of Janus 1T-MoSH (8.38~31.95 N·m⁻¹, see **Figure S1(a)**) and 1T'-MoSH (7.37~19.49 N·m⁻¹, see **Figure S1(b)**), while significantly lower than that of 1T'-MoSF (4~132 N·m⁻¹)[41].

**Figure 2(b)** reveals the angular-dependent distribution of $v(\theta)$, demonstrating how the transverse-axial strain relationship changes with crystallographic directions. This is critical for analyzing lateral deformation under external loading. **Figure 2(b)** and **Figure S1** demonstrate that the $v(\theta)$ values (0.47~0.97) fall within a comparable range to those of 1T-MoSH (0.68~0.92) and 1T'-MoSH (0.64~0.87), with overlapping numerical distributions. **Figure 2 (c)** presents a surface plot of strain energy ($\Delta E$) versus in-plane strains ($\varepsilon_x$, $\varepsilon_y$). It quantitatively depicts energy changes under diverse in-plane strain combinations, offering insights into the Hybrid 1T'-MoSH's energy dissipation and storage mechanisms during deformation. The prominently elliptical shape of $\Delta E$ indicates the mechanical anisotropy of the Hybrid 1T'-MoSH. Under uniaxial strain, tensile strains ($\varepsilon_x = 0\%~5\%$) exhibit $\Delta E > 1.5$ eV/cell, while compressive strains ($\varepsilon_x = -5\%~0\%$) show $\Delta E < 1.5$ eV/cell. Similar results are also observed for the strains of $\varepsilon_y$. These results demonstrate that the strain energy demanded for equal-magnitude tensile deformation is markedly lower than that for compressive deformation. Moreover, for the same degree of biaxial strain, the energy required to apply strain in the same direction along both axes is greater than that for applying strain in different directions along the two axes. This finding reveals that the Hybrid 1T'-MoSH exhibits a positive $v(\theta)$.

As shown in **Figure 3 (a)**, we also performed the *ab* initio molecular dynamics for Hybrid 1T'-MoSH monolayers in a 4×4 supercell. For temperatures ranging from 300 K to 600 K, the energies of Hybrid 1T'-MoSH monolayers exhibit stable oscillations within 5 ps. This observation strongly indicates the material's exceptional thermodynamic stability in this temperature regime. Upon elevating the temperature to the 900 K~1500 K range, although the energies exhibit substantial fluctuations within 5 ps, the structural integrity of Hybrid 1T'-MoSH monolayers remains intact. Notably, upon raising the temperature to 1600 K, the energy shows pronounced oscillations within 5 ps, and hydrogen atoms in the Hybrid 1T'-MoSH monolayers start to escape from the interface, forming $H_2$ (**see the inset in Figure 3(a)**). This finding clearly demonstrates that the maximum temperature threshold that Hybrid 1T'-MoSH monolayers can withstand is below 1600 K. Subsequently, a dynamic stability analysis of the Hybrid 1T'-MoSH structure was carried out. Density functional perturbation theory was employed to compute the phonon dispersion of Hybrid 1T'-MoSH monolayers along the Γ - X - Y - Γ path within the Brillouin zone. As depicted in **Figure 3 (b)**, 18 phonon branches are present. In the low-frequency regime (near 0 cm$^{-1}$), three of these are acoustic branches, corresponding to the longitudinal acoustic (LA), transverse acoustic (TA), and zonal acoustic (ZA) modes, respectively. Notably, the phonon spectrum exhibits no imaginary frequencies, signifying the dynamic stability of the Hybrid 1T'-MoSH structure.

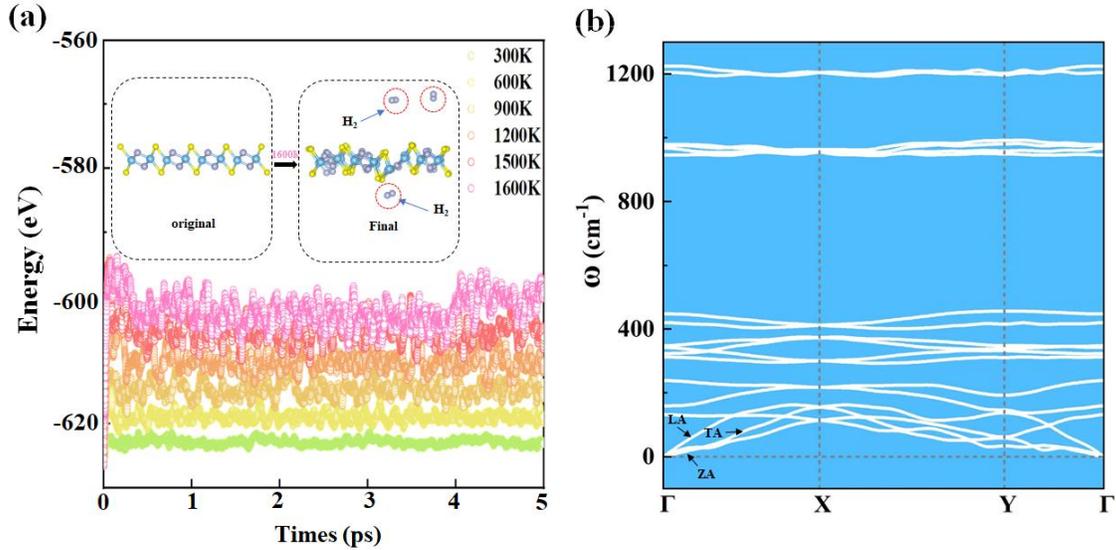

**Figure 3.** (a) Thermodynamic stability analysis conducted using the AIMD simulations at 300 K~1500K. The display snapshots of atomic structures for the Hybrid 1T'-MoSH monolayers at 0 ps and 5 ps from a top views. (b) The phonon dispersion of Hybrid 1T'-MoSH monolayers. Notion: transverse acoustic (TA), longitudinal acoustic (LA), zonal acoustic (ZA).

To investigate the Raman activity of Hybrid 1T'-MoSH monolayer, we computed the Raman spectra. As shown in **Figure 4(a)**, five distinct Raman peaks are presented, indicating that five optical-phonon modes are Raman-active in the Hybrid 1T'-MoSH monolayers. Among them, three modes (I, II, III) are in the low-frequency region, and two modes (IV, V) are in the high-frequency region. Further, we analyzed the atomic vibration modes corresponding to the Raman peaks, as depicted in **Figure 4(b)**. The modes of Ⅰ and Ⅲ show the phonon vibration patterns of the Raman peaks at 136.7 cm$^{-1}$ and 354.4 cm$^{-1}$, respectively. It can be seen that the Raman peaks at these two positions are both induced by the in-plane relative vibrations of Mo and S atoms. At 136.7 cm$^{-1}$, the vibrational directions of the upper-layer S atoms are opposite to those of the lower-layer S atoms. Conversely, the upper- and lower-layer S atoms vibrate in the same direction at 354.4 cm$^{-1}$. The figure corresponding to mode II shows the phonon vibration pattern of the Raman peak at 259.2 cm$^{-1}$, which is dominated by the coupling of in-plane and out-of-plane vibrations of Mo and S atoms, with a negligible contribution from H atoms. Notably, the Raman peak induced by the phonon vibration at 964.4 cm$^{-1}$ (IV) is the most active. This is caused by the intense out-of-plane vibration of H atoms. This vibration mode has a relatively high energy and is mainly dominated by the rapid stretching of Mo-H bonds. For the atomic vibration at 1288.9 cm$^{-1}$ (V), the counter-vibration of S atoms contributes the most to the Raman peak at this position. Here, the S atoms, affected by the high-frequency vibration of the Mo-H skeleton, also perform high-frequency and small-amplitude vibrations near specific positions, overall exhibiting the characteristics of high-frequency vibrations under strong interactions. These results suggest that in the lower-frequency optical branches, the Raman peaks are mainly caused by the vibrations of Mo and S atoms, with little contribution from H atoms. In contrast, in the higher-frequency optical branches, the intense vibrations of H atoms mainly give rise to the Raman peaks, and Mo atoms do not participate in the vibrations.

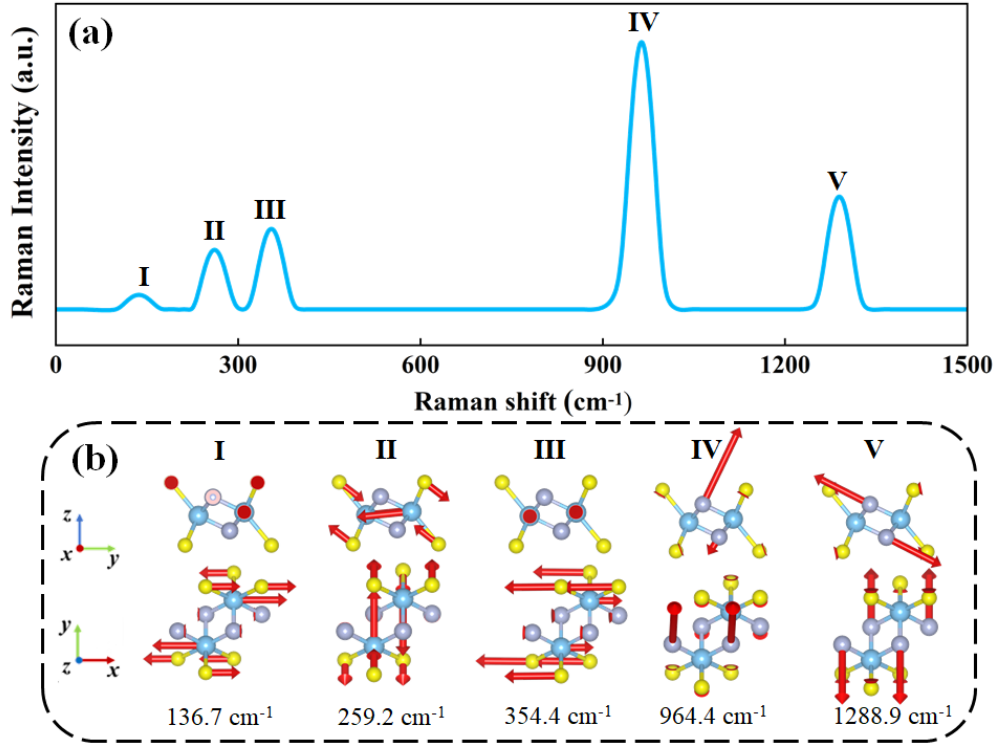

**Figure 4.** (a) Raman spectra of the Hybrid 1T'-MoSH, (b) The vibrational characteristics of the Raman-active phonon modes. The vibration mode diagram was drawn using the VESTA software[42].

**Figure 5 (a)** presents the orbital-resolved band structures and projected density of states (PDOS) of Hybrid 1T'-MoSH, with the Fermi level set at $E_f=0$. The band structure reveals that the conduction band (CB) and valence band (VB) cross the Fermi level, indicating the metallic nature of Hybrid 1T'-MoSH. This finding is similar to the band structures observed in Janus 2H-MoSH, 1T-MoSH, and 1T'-MoSH[35]. From the PDOS, we observe that the electron contributions from the Mo atom's $d$ orbitals dominate the electronic states at the Fermi level. In contrast, the contributions from the S atom's $p$ orbitals are relatively minor, and those from the Mo atom's $s$ orbitals are negligible. Using group-theory concepts, the Mo atom's $d$ orbitals can be classified into three types based on the point group $A'_1(d_z^2)$, $E'_1(d_{xy,x^2-y^2})$, $E_1(d_{xz,yz})$, where $A'_1$, $E'_1$ and $E_1$ are the Mulliken symbols of the irreducible representations[37]. In the orbital-resolved band structures, around the high-symmetry point

Γ, the bands are relatively flat and mainly governed by the $d_{xy,x^2-y^2}$ orbitals. Near the X point, the electronic properties are mainly associated with the hybridization of $d_{xy,x^2-y^2}$ and $d_{xz,yz}$ orbitals. Away from the high-symmetry points and the Brillouin-zone boundaries, the mixing of various Mo atom orbitals contributes to the band structure. As shown in **Figure 5(b)**, the Fermi surface of Hybrid 1T'-MoSH has two parts as two bands cross the Fermi surface. The Fermi sheets near the Γ point originate from the hybridization of $d_{xy,x^2-y^2}$ and $d_{xz,yz}$ orbitals, while the outer Fermi sheet is related to the $d_{xy,x^2-y^2}$ orbitals.

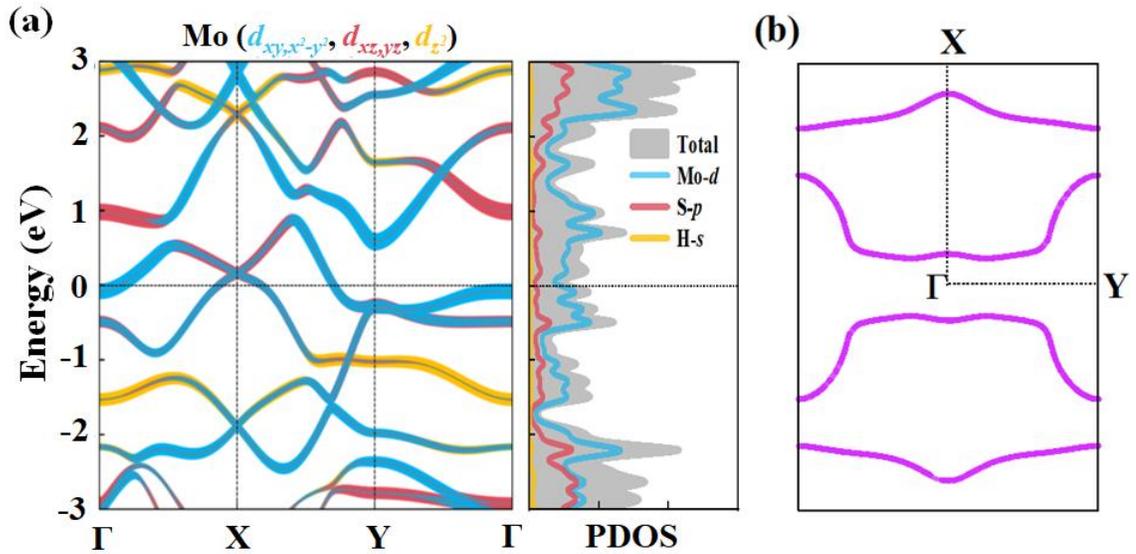

**Figure 5.** (a) The orbital-resolved band structures and density of states of the Hybrid 1T'-MoSH. The blue, red, and yellow lines highlight the $d_{xy,x^2-y^2}$, $d_{xz,yz}$, and $d_{z^2}$ orbitals of the Mo atom, respectively. The Fermi level is set to zero. (b) Fermi surface.

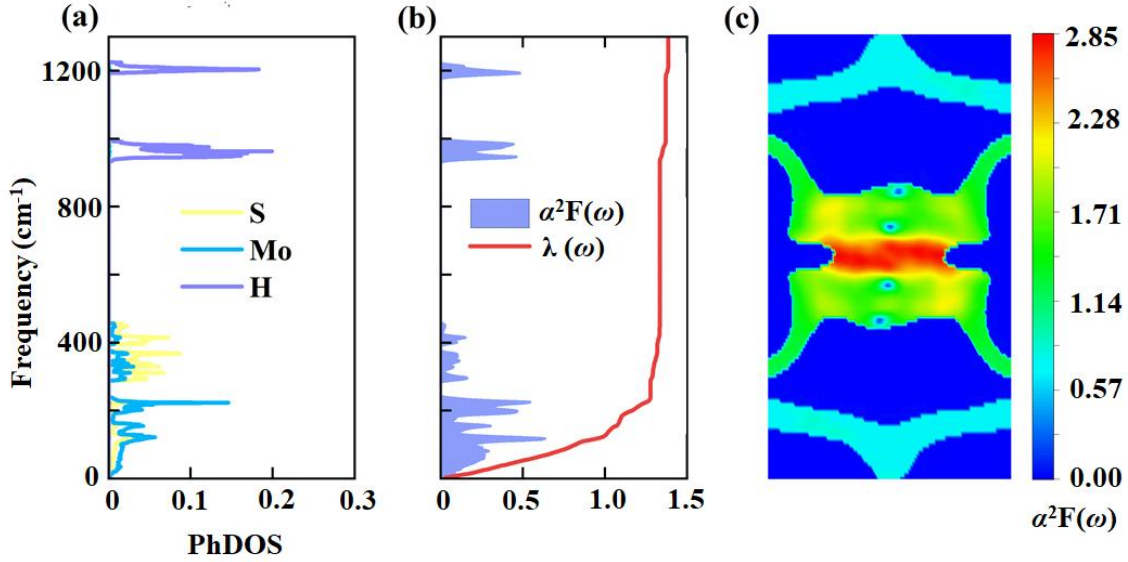

**Figure 6.** (a) Phonon density of states (PhDOS) for Hybrid 1T'-MoSH, (b) The Eliashberg spectral function α²F(ω) and the cumulative contribution to the Electron-Phonon Coupling (EPC) strength λ(ω), (c) The distribution of EPC on the Fermi surface.

Subsequently, we further analyzed the vibrational characteristics and Electron-Phonon Coupling (EPC) of the Hybrid 1T'-MoSH monolayer. As shown in **Figure 6**, from left to right, the figure presents the phonon density of states (PhDOS), the Eliashberg spectral function $\alpha^2F(\omega)$, the integrated EPC constant $\lambda(\omega)$, and the distribution of EPC on the Fermi surface. Based on the contribution distribution of different atoms to the PHDOS in **Figure 6(a)**, the vibration of the H atoms occupies the high-frequency region, around 900 cm⁻¹ to 1200 cm⁻¹. This is because the phonon vibration frequency is inversely proportional to the square root of the element mass[39]. Then, the S atoms mainly occupy the mid-frequency range from 300 cm⁻¹ to 500 cm⁻¹, while the Mo atoms are mainly located around 210 cm⁻¹. These distributions are consistent with the Raman spectra in **Figure 4(a)**. To analyze the coupling strength between different phonon modes and electrons, we combined $\alpha^2F(\omega)$ and $\lambda(\omega)$ to calculate the total EPC. As shown in **Figure 6(b)**, the total EPC strength obtained through integration is 1.39 (red line), which is lower than that of the previously studied

Janus 1T - MoSH (2.37) phase[39] and Janus 2H - MoSH (1.48) phase[37]. To quantify the anisotropy of EPC, we further study the momentum-resolved EPC parameter $\lambda_{nk}$. The variation of the EPC parameter $\lambda_{nk}$ over Fermi surface (FS) is shown in **Figure 6(c)**. The region of the EPC strength is strongest along the Γ-Y path and weakest at the Brillouin zone boundary, mainly associated with states in the Mo $d$ orbitals. The results of this analysis are consistent with the previously studied 2H-MoSH[37].

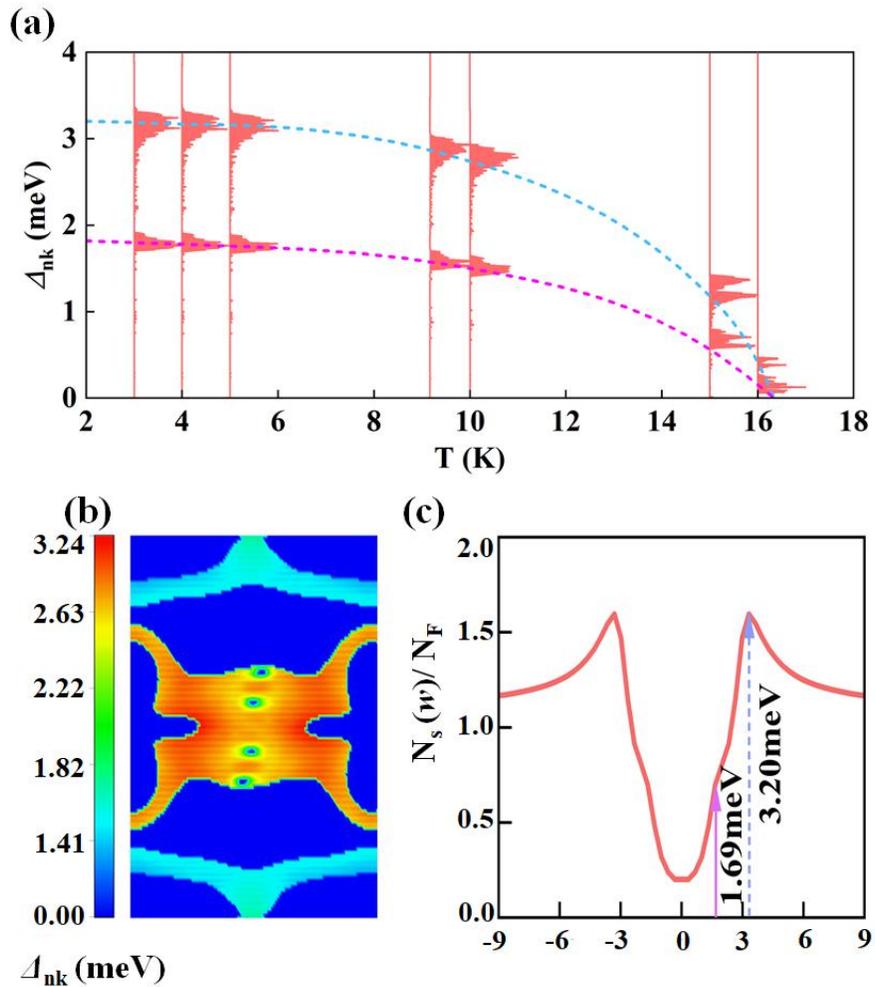

**Figure 7.** (a) Calculated anisotropic superconducting gaps of Hybrid 1T'-MoSH on the Fermi surface as a function of temperature，（b）The distribution of the superconducting gaps on the FS at 10 K，（c）The normalized superconducting density of states (SDOS) at 10 K.

We analyze the superconducting properties of Hybrid 1T'-MoSH using momentum-resolved gaps $\Delta_{nk}$ ($\omega=0$), normalized superconducting density of states (SDOS), and the temperature-dependent distribution of superconducting gaps $\Delta_{nk}$ ($\omega=0$) on the Fermi surface. To accurately calculate $\Delta_{nk}$ and the superconducting transition temperature $T_c$, we apply the anisotropic Migdal - Eliashberg theory. This theory is based on electron - phonon Wannier - Fourier interpolation in the EPW code, with a Coulomb pseudo-potential $\mu^*=0.1$. As depicted in **Figure 7(a)**, the superconducting gaps exhibit a standard BCS-type temperature dependence. As the temperature rises, the superconducting gap diminishes until it vanishes at the superconducting transition temperature $T_c$. At each temperature, the gaps cluster around two distinct values, demonstrating significant anisotropy in the gap energy spread and indicating that Hybrid 1T'-MoSH is a two-gap superconductor. Ultimately, we determine the superconducting transition temperature $T_c$ of Hybrid 1T'-MoSH to be 16.34 K, and its maximum zero-temperature superconducting gap to be 3.21 meV. These values are lower than the $T_c$ of 28.58 K and the maximum zero-temperature superconducting gap of 6 meV for Janus 2H-MoSH[37], respectively. To further analyze To further investigate the origin of $\Delta_{nk}$, we calculated and plotted the distribution of the momentum-resolved superconducting gap $\Delta_{nk}$ on the Fermi surface at 10 K, as shown in **Figure 7(b)**. The distribution of the superconducting gap on the Fermi surface is similar to that of $\lambda_{nk}$. The maximum superconducting gap opened near the $\Gamma$ point has an energy of approximately 3.24 meV, which is mainly related to the orbital hybridization of $d_{xy,x^2-y^2}$ and $d_{xz,yz}$. The minimum superconducting gap opened on the Fermi sheet at the Brillouin zone boundary is approximately 1.83 meV, mainly originating from the $d_{xy,x^2-y^2}$ orbitals.

**Figure 7(c)** shows the normalized superconducting density of states (SDOS) of Hybrid 1T'-MoSH at 10 K. To visually demonstrate the two-gap superconductivity of MoSH, the SDOS is calculated by using the Padé approximation to solve the anisotropic Migdal-Eliashberg equation for the isotropic gaps $\Delta_{nk}$ on the real axis. The formula for calculating

the SDOS is

$$\frac{N_s(\omega)}{N_F(E_F)} = \text{Re}[\frac{\omega}{\sqrt{\omega^2 - \Delta^2(\omega)}}]$$

Where $N_S(\omega)$ and $N_F(E_F)$ are respective DOS of superconducting state and normal state at the Fermi level. The vertical arrows of the blue and purple dashed lines indicate the average values of the superconducting gaps induced by the *d*-orbital electrons of Mo atoms at 10 K. These values are related to the central peaks in the distribution of superconducting gaps as a function of temperature. Notably, the two peaks in the SDOS provide strong evidence for the two- gap superconductivity of Hybrid 1T'-MoSH.

To further validate the effectiveness and extendibility of our structural model, we replaced the Mo atoms in Hybrid 1T'-MoSH with Hf, Ta, and Ti atoms respectively, forming Hybrid 1T'-HfSH, TaSH, and TiSH monolayers. The structures are shown in **Figure S2**. Similar to Hybrid 1T'-MoSH, the orbital - resolved band structures and density of states in **Figure S3** indicate that the Hybrid 1T'-HfSH, TaSH, and TiSH monolayers also exhibit metallic characteristics, and the *d* orbitals of the metal atoms contribute the most near the Fermi level. As shown in **Figure S4**, we calculated the phonon spectra, phonon density of states, and EPC strengths of these three structures. The results show that these three structures are dynamically stable. Although there are small imaginary frequencies at the Γ point, they can be eliminated by increasing the sampling of *K* points in the Brillouin zone. Similar to Hybrid 1T'-MoSH, the vibrations of phonons in the high-frequency region are mainly contributed by H atoms, while those in the low-requency region are mainly contributed by the metal atoms Hf, Ta, and Ti. As shown in **Table 2**, the EPC of HfSH, TiSH, and TaSH are 0.38, 0.46, and 0.57, respectively. These values are much lower than 1.39 of Hybrid 1T'-MoSH, suggesting that their $T_c$ will be lower than that of Hybrid 1T'-MoSH. We also calculated the $T_c$ of HfSH, TiSH and TaSH monolayers. The results show that HfSH, TiSH, and TaSH monolayers have transition temperatures of 0.53 K, 1.59 K, and 2.42 K respectively, which are much lower than 16.34 K of Hybrid 1T'-MoSH. These results underscore the unique role of Mo in enhancing the superconductivity of Hybrid 1T'-MoSH.

**Table 2.** EPC strengths, superconducting transition temperature ($T_c$) of HfSH, TiSH, TaSH, and Hybrid 1T'-MoSH monolayer, in GPa.

| Systems | EPC | $T_c$ |
| --- | --- | --- |
| **HfSH** | 0.38 | 0.53 |
| **TiSH** | 0.46 | 1.59 |
| **TaSH** | 0.57 | 2.42 |
| **Hybrid 1T'-MoSH** | 1.39 | 16.34 |

## 3. Methodology

Using the Quantum ESPRESSO (QE) package[43-44] that bases on density functional theory, we conducted the first-Principles calculations. The norm-conserving pseudopotentials (NCPPs)[45] were employed to describe the atomic core. The local density approximation[46] was adopted for the exchange-correlation functional, along with a plane wave cutoff set at 80 Ry. For structural optimization, the convergence criteria for energy and force are $10^{-6}$ Ry and $10^{-7}$ Ry/Bohr, respectively. In the electronic structure and density of states (DOSs) calculations, Monkhorst–Pack sampling grids with a resolution of $2\pi \times 0.01$/Å. The phonon dispersions are calculated within density functional perturbation theory (DFPT) on a $12 \times 12$ q-mesh. To investigations into superconducting properties entailed solving the anisotropic Migdal–Eliashberg equation within a denser $120 \times 120 \times 1$ k-grid and $60 \times 60 \times 1$ q-grid by means of the EPW code[47]. To ensure a reliable and precise calculation of the superconducting transition temperature $T_c$, the band structure was interpolated via the Wannier function. A vacuum layer with a minimum thickness of 15 Å was introduced to minimize the interactions caused by periodic structures.

The superconducting properties of Hybrid 1T'-MoSH is investigated through the EPW code. We solve fully anisotropic Migdal-Eliashberg equations through an imaginary axis at the fermion Matsubara frequencies $\omega_j$ (*j* is an integer) for a series of temperatures with

the same interval according to[47-48]:

$$Z_{nk}(i\omega_j) = 1 + \frac{\pi T}{N_F \omega_j} \sum_{mk'j'} \frac{\omega_{j'}}{\sqrt{\omega_{j'}^2 + \Delta_{mk'}^2(i\omega_{j'})}} \times \delta(\epsilon_{mk'}) \lambda(nk, mk', \omega_j - \omega_{j'})$$

$$Z_{nk}(i\omega_j)\Delta_{nk}(i\omega_j) = \frac{\pi T}{N_F} \sum_{mk'j'} \frac{\Delta_{mk'}(i\omega_{j'})}{\sqrt{\omega_{j'}^2 + \Delta_{mk'}^2(i\omega_{j'})}} \times \delta(\epsilon_{mk'}) \left[\lambda(nk, mk', \omega_j - \omega_{j'}) - \mu_c^*\right]$$

where $Z_{nk}(i\omega_j)$ is the mass renormalization function, $\Delta_{nk}(i\omega_j)$ is the superconducting gap function, $N_F$ is the electronic density of states (DOS) per spin at the Fermi level, $\epsilon_{mk'}$ is the energy eigenvalue of the electronic state $|mk'>$, $\mu^*_c$ is the effective screened Coulomb repulsion constant, as discussed below, we use the widely accepted value of $\mu$=0.1 to evaluate the $T_c$. $\lambda(nk, mk', \omega_j - \omega_{j'})$ represents the momentum- and energy dependent EPC parameter, which is written as:

$$\lambda(nk, mk', \omega_j - \omega_{j'}) = N_F \sum_v \frac{2\omega_{qv}}{(\omega_j - \omega_{j'})^2 + \omega_{qv}^2} |g^v_{nk,mk'}|^2 = 2\int_0^\infty \frac{\alpha^2 F(\omega)}{\omega} d\omega$$

where $g^v_{nk,mk'}$ denotes the electron-phonon matrix element arising from the scattering between the states $\epsilon_{mk'}$ and $\epsilon_{mk}$ via the phonon mode $\epsilon_{qv}$ with wave vector q $=k - k'$ and frequency $\omega_{qv}$.

## 4. Conclusion

In this study, we successfully identified a novel non-Janus Hybrid 1T'-MoSH monolayer through a combination of high-throughput screening and first-principles calculations. This structure represents a hybridization of MoS$_2$ and MoH$_2$, with unique atomic arrangements that endow it with excellent stability in terms of energy, mechanics, dynamics, and thermodynamics. Specifically, it has the lowest binding energy of - 3.02 eV among 896 predicted structures, indicating high energy stability. The mechanical stability satisfies the lattice stability criteria, and it shows good dynamic stability as evidenced by the absence of imaginary frequencies in the phonon spectrum. Thermodynamically, it remains stable below 1600 K. In terms of electronic and superconducting properties, the Hybrid 1T'-MoSH monolayer is metallic, with Mo *d* orbitals dominating the electronic states at the Fermi level. It is predicted to be a two-gap superconductor with a superconducting transition temperature $T_c$ of 16.34 K. The substitution of Mo with Hf, Ta, or Ti in Hybrid 1T'-MoSH leads to significantly lower $T_c$ values (0.53~2.42 K), highlighting the crucial role of Mo in enhancing superconductivity. This work designs a new 2D transition metal chalcogenide material---the hybrid 1T'-MoSH monolayer, which enriches the family of 2D materials and provides a promising candidate for future quantum devices.

## Supporting Information

Supporting Information is available from the Wiley Online Library or from the author.

## Acknowledgements

This research was supported by the National Natural Science Foundation of China under Grants No. 12204402, the School-level Cultivation Project of the National Natural Science Foundation of China in 2024, No. RZ2400003824, Talent Special Project, No. DC2400003207, the Natural Science Foundation of Guangxi Province (No. 2022GXNSFAA035487), and the Hefei Advanced Computing Center.

## Conflict of Interest

The authors declare no conflict of interest.

## Data Availability Statement

The data that support the findings of this study are available from the corresponding author upon reasonable request.

## Keywords

MoSH monolayer, Non-Janus structure, High-throughput screening, Superconductivity